\documentclass[num-refs]{wiley-article}

\usepackage{siunitx}
\usepackage{float}
\usepackage{amsmath}
\graphicspath{ {./image/} }
\papertype{Original Article}

\title{Intraday trading strategy based on time series and machine learning for Chinese stock market}

\author[1]{Qinan Wang}
\author[1]{Yaomu Zhou}
\author[1]{Junhao Shen}

\affil[1]{Lyle School of Engineering, Southern Methodist University, Dallas, TX, 75275, USA}

\corraddress{Qinan Wang, M.S.\\ Lyle School of Engineering, Southern Methodist University, Dallas, TX, 75275, USA}
\corremail{qinanw@smu.edu}

\runningauthor{Wang et al.}

\begin{document}

\begin{frontmatter}
\maketitle

\begin{abstract}
This article comes up with an intraday trading strategy under T+1 using Markowitz optimization and Multilayer Perceptron (MLP) with published stock data obtained from the Shenzhen Stock Exchange and Shanghai Stock Exchange. The empirical results reveal the profitability of Markowitz portfolio optimization and validate the intraday stock price prediction using MLP. The findings further combine the Mark- owitz optimization, an MLP with the trading strategy, to clarify this strategy's feasibility.
\end{abstract}
\end{frontmatter}


\section{Introduction} 
The development of The Chinese stock market (A shares) since its inception has been impressive. Especially in recent years, the Chinese stock market has been strong, which seems to sign a bull market. Most of these companies have outperformed expectations and performed well. Except for a few losers, all sectors have performed strongly. It raises attention worldwide. We find that it is invaluable to research quantitative methods for the stock market under this condition.\\\\
In China's stock market, the regulator implements T+1. That is, stocks can only be sold the next day after they are bought on the same day. It is hard to figure out a way of daily trading in the market. However, the price of a stock fluctuates every day. The maximum amplitude of a single stock can even reach 20\% (the daily cap and collar volatility difference: upper limit is 10\%, lower limit is -10\%). Given this circumstance, we can carry out the following operations: Open positions in one day advance. On the following day, sell the stocks at a high intraday price, buy the same quantity of positions as selling before at a low price of the day; Or buy at a low price and sell at a high price. At this point, we have managed a similar T+0 operation. We make a profit by the difference in price within the day. During this process, the stock account share will not change, but the amount of money available in the account will increase.\\\\
In previous publications, the stock price prediction has been researched in many models. Tang et al. compared their model of autoregressive moving average generalized autoregressive conditional heteroscedasticity (ARMA-GARCH) with the finite mixture of autoregressive (AR), a finite mixture of the autoregressive moving average (ARMA) and a finite mixture of autoregressive generalized autoregressive conditional heteroscedasticity (AR-GARCH) models for finance exchange rate prediction. The experimental simulation indicated that the mixture of the ARMA-GARCH model derives a GEM algorithm with better performance than others. \cite{tang2003finite} In 2007, Ince et al. utilized technical indicators with heuristic models, kernel principal component analysis, and factor analysis to select the most useful inputs for a predicting model. They used different inputs on multilayer perceptron (MLP) networks and support vector regression (SVR). SVR and MLP networks required those different inputs studied on comparison studies. Furthermore, proposed heuristic models produced better results than the data mining method. Besides, there was no difference between MLP and SVR techniques on their mean square error values. \cite{ince2007kernel} Gong et al. created a new method with Logistic Regression to forecast the next month's stock price trend basing on the current month in 2009. They chose Shenzhen Development stock A(SDSA) from RESSET Financial Research Database as a study case. Comparing with other models, e.g., RBF-ANN forecasting model, their model was less complicated and at least 83\% accurate in prediction. \cite{gong2009new} In 2010, Budiman et al. used the Artificial Neural Network (ANN) and ARIMA method to predict the results of ANTM (PT. Aneka Tambang) of stock. The study showed that using the ANN method had a smaller error than the ARIMA approach. \cite{wijaya2010stock}  After three years, Alkhatib et al. forecasted stock price for a sample of six companies on the Jordanian stock exchange utilizing the K- nearest neighbor algorithm and nonlinear regression method to help investors, management, decision-maker, and users make an accurate decision. Basing on the predicting result, which is rational and reasonable, the KNN method was excellent with the smallest error. Compared with the real stock price data, the forecasting result and the actual stock price were almost parallel. \cite{alkhatib2013stock} In 2016, Persio et al. predicted stock price by using the Artificial Neural Network approach. They also considered MLP, CNN, LSTM recurrent neural networks techniques, and S\&P500 historical time series to predict the stock price trend. As a result, they indicated that neural networks could forecast future movements' financial time series and propose more ways to improve the results. \cite{di2016artificial} In recent years, there is some research on predicting stock price by deep learning method with algorithm development. For example, in 2019, Nikou et al. studied to assess the forecasting power by Machine Learning models in a stock market. They created four MLA models to make predictions. The results showed that the deep learning approach was best than others. The second-best method was supports vector regression concerning neural network and random forest methods with less error. \cite{nikou2019stock}
\\\\
Most of the researches are only for price prediction since the T+1 regulation of China's stock market makes intraday trading difficult. It is innovative to research the intraday trading strategy for China's stock market. Besides, Guresen et al. provided a detailed survey for 25 articles of the NNs application on the finance and applied a multi-layer perceptron model, a dynamic artificial neural network and hybrid networks to predict NASDAQ Stock Exchange Index. They concluded that NN-based solutions outperform other statistical approaches in most cases. \cite{guresen2011using} Hence, we combine the prediction by multilayer perceptron (MLP) and practical operation to design the strategy. This strategy is validated by simulated trade.
\\\\
In this strategy, to assure both the profit of origin holdings and intraday trading, we select a portfolio consists of promising stocks, and manage risk. To avoid unexpected volatility of certain industries or companies, we get three stocks from 10 industries, respectively. These industries are Science and Technology, Basis Materials, Consumption (periodic), Finance, Service, Means of Production, Energy, Consumption (aperiodic), Health Care, transportation. In each industry, we use the top three performance stocks of the last month.On the first day of trading, we open positions for all stocks in the portfolio. On the following days, we trade two times each day to gain from the intraday volatility. The daily trading hours for continuous bidding of A shares are from 9:30 to 11:30 in the morning and from 13:00 to 15:00 in the afternoon. The trading hours are only 4 hours a day, and there is a 1.5-hour break at noon. We use one-minute trading data from 9:30 to 11:20 to train the multilayer perceptron (MLP). This model is used to predict the stock price of 13:00 to 14:50. With the price point at 14:50, we make trading decision by the difference between this point and price at 11:20. If the prediction price is lower than the price at 11:20, we sell holding positions before the close in the morning and buy back at 14:50. If the prediction price is higher than the price at 11:20, we buy additional positions before the close in the morning and sell them at 14:50. The model is retrained every day for a possible large gap between successive days. The goal of this article is to gain profit by intraday trend prediction. \\\\
In the next section, we describe the data we used in the analyses. The Method section contains Markowitz portfolio optimization, multi-layer perceptron (MLP), daily volatility, and measurement of return. The Results section includes the results of portfolio optimization, modelling, and an evaluation of the return of the strategy. The last section discusses the model performance, further improvement, and comparison with other models.\\\\

\section{Data}
\begin{figure}[h]
\centering
\includegraphics[scale=0.13]{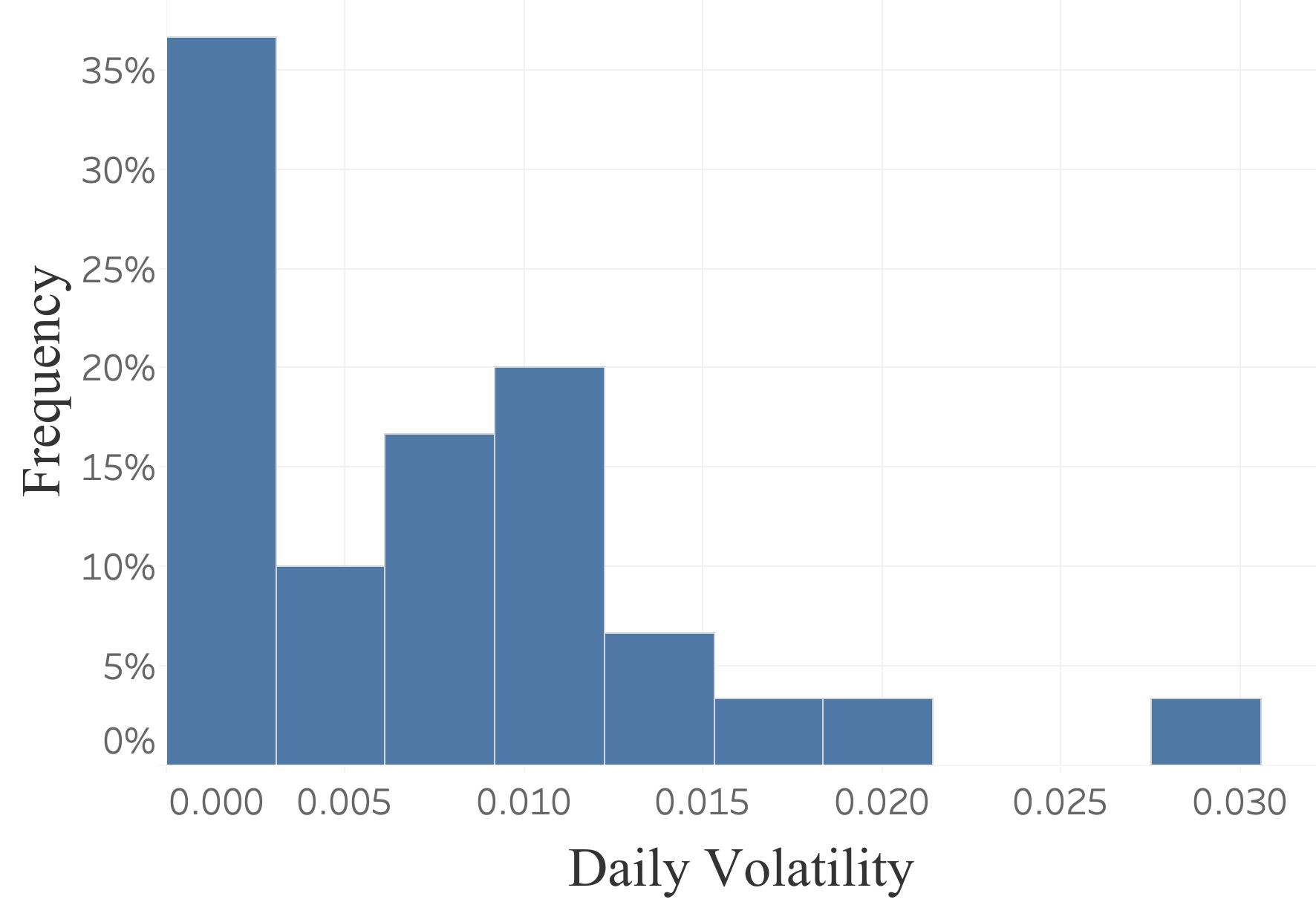}
\caption{Histogram of daily volatility}
\label{fig:pmf}
\end{figure}
\noindent Given the previous performance in 2020, we pick 30 stocks for the portfolio. This study's experimental data include close price of minutes data in October and close price of daily data from January 2020 to September 2020. We collected the data from Xueqiu Finance. Daily data is used in Markowitz Optimization. Minutes data is used for day trade. The data plots are as Figure \ref{fig:minute_data} and Figure \ref{fig:daily_data} (See Appendix). The tickers information is as Table \ref{table:ticker} (See Appendix). In Figure \ref{fig:pmf}, the daily volatility ranges from 0 to 0.03. We have a 0.6 probability that the daily volatility is over 0.002. It shows a potential earning for the day trade strategy.
\section{Method}
\subsection{Markowitz portfolio optimization}\label{sec:var}
\subsubsection{Model}
Markowitz's portfolio theory was invented in 1952. He proposed the portfolio optimization model using the mean and variance of individual stock returns in an asset portfolio to find the Efficient Frontier of an investment portfolio, the portfolio with the lowest variance under a certain yield level. According to Markowitz's portfolio model, it is necessary to invest in diverse stocks and ensure that stocks' correlation coefficient is low to minimize a portfolio's risk. From the relationship between the return and risk of risky assets, it discusses selecting the economic system's optimal asset portfolio. The portfolio selection model focuses on risk diversification and quantitatively determine the portfolio.\\\\
The apparent form of parametric optimization of Markowitz model has the following mathematical representation:\\\\
\begin{equation*}
\begin{array}{l}
\max \mathbb{E}\left(r_{P}\right)=\max \sum_{i=1}^{n} \omega_{i} \mu_{i} \\
\min \sigma_{P}=\min \sqrt{\sum_{i=1}^{n} \sum_{j=1}^{n} \omega_{i} \omega_{j} \sigma_{i j}} \\
0 \leq \omega_{i} \leq 1, \quad i=1, \ldots, n \\
\sum_{i=1}^{n} \omega_{i}=1
\end{array}
\end{equation*}
\\\\
where $\omega_{i}$ is the percentage of capital that will be invested in asset $i ; r^{i}$ is the return on asset $i ; \mu_{i}$ the expected return on asset $i ; \mu_{i j}$ is the covariance between the return on assets $i$ and $j ; \mathbb{E}\left(r_{P}\right)$ is the expected return of the portfolio; $\sigma_{P}$ is the risk of the portfolio.\cite{markowitz2000mean}
The Markowitz model is based on several assumptions concerning the behavior of investors and financial markets:\cite{chen2010portfolio}\\\\
\vspace{-\topsep}
\begin{itemize}
  \setlength{\parskip}{0pt}
  \setlength{\itemsep}{0pt plus 1pt}
    \item A probability distribution of possible returns over some holding period can be estimated by investors.
    \item Investors have single-period utility functions in which they maximize utility within the framework of diminishing marginal utility of wealth.
    \item Variability about the possible values of return is used by investors to measure risk.
    \item Investors care only about the means and variance of the returns of their portfolios over a particular period.
    \item Expected return and risk as used by investors are measured by the first two moments of the probability distribution of returns-expected value and variance.
    \item Return is desirable; risk is to be avoided.
    \item Financial markets are frictionless.
\end{itemize}
\vspace{-\topsep}
\subsubsection{Sharpe ratio}
The Sharpe ratio can be used to measure the return on investment at a given risk. This ratio adjusts the return on investment, allowing us to compare different investments' performance at a given scale of risk. Without the limitation of scale risk, we cannot compare the returns and risk performance of different securities portfolios.\\\\
\begin{equation*}
\text { Sharpe Ratio }=\frac{R_{p}-R_{r f}}{\sigma_{p}} 
\end{equation*}\\\\
where $R_p$ is the expected portfolio/asset return, $R_{rf}$ is the risk-free rate of return, $\sigma_{p}$ is portfolio/asset standard deviation.\\\\

\subsection{Multilayer Perceptron (MLP)}
\begin{figure}[h]
\centering
\includegraphics[scale=0.22]{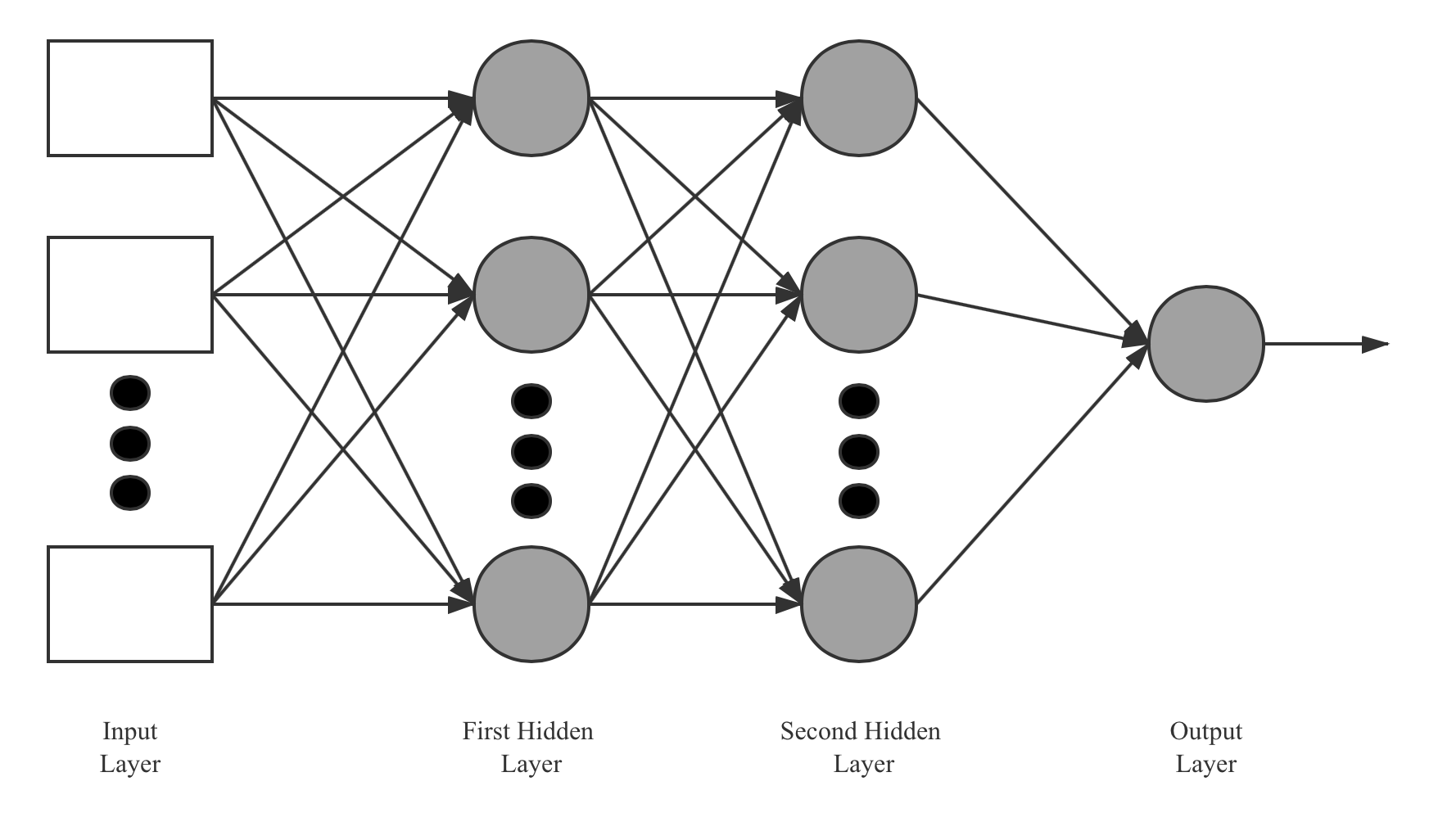}
\caption{Multilayer Perceptron structure}
\label{fig:mlp}
\end{figure}

The architecture of artificial neural networks (ANNs) is based on connections of layers by nodes called neurons as well as the biological neurons of brain. \cite{ecer2013artificial} Each path transmits a signal among neurons in a manner similar to that of synapses. \cite{ardabili2016modeling} MLP, as a feedforward ANN, contains three main parts: one input layer, one or more hidden layers and one output layer, which can be successfully employed for prediction, classification, signal processing and error filtering. \cite{ecer2013comparing} Each node employs one nonlinear function. MLP employs backpropagation learning algorithm for training process. \cite{gundoshmian2019prediction,ardabili2014simulation} MLP as popular and frequently used techniques among other MLPs was employed to predict the direction value.  Figure \ref{fig:mlp} indicates the architecture of developed network. \cite{ecer2020training}
\subsection{Daily volatility}
To verify the strategy feasibility,  we need the probability mass function (pmf) of the volatility. Instead of represented by the standard deviation, the volatility is defined as the formula: \\\\
\begin{equation*}
    \sigma = \frac{|p_a - p_b|}{p_a} 
\end{equation*}\\\\
where $p_a$ is the price of the stock at 11:20, $p_{b}$ is the price of the stock at 14:50, $\sigma$ is the daily volatility.\\\\
\subsection{Measurement of return}
\subsubsection{Monthly return}
In this strategy, we will have origin positions for 30 stocks. The monthly return for each stock is defined as the formula:\\\\
\begin{equation*}
    r_m = \frac{p_e - p_s}{p_s}
\end{equation*}\\\\
where $r_{m}$ is the monthly return of the stock; $p_{e}$ is the selling price; $p_{s}$ is the buying price.
\subsubsection{Daily return}
In this strategy, we will trade a stock two times per day, selling in the morning and buying in the afternoon, vice versa. We assume that trading the same amount as origin position each day. Given any stock and its daily buying, daily selling price, origin position price, we have the daily return of the stock as the formula:\\\\
\begin{equation*}
    r_t = \frac{s_t - b_t}{o}
\end{equation*}\\\\
where $r_{t}$ is the return of the stock at time $t$, $s_{t}$ is the selling price at time $t$, $b_{t}$ is the buying price of the stock at time $t$, $o$ is the origin position price.
\subsubsection{Yearly return}
Given any stock and its daily or monthly return, the formulas of expected yearly return for this stock are:
\begin{equation*}
    \mathbb{E}\left(r_{y}\right) = (1 +  \mathbb{E}\left(r_{m}\right))^{12} - 1
\end{equation*}
\begin{equation*}
    \mathbb{E}\left(r_{y}\right) = (1 + \sum_{t=1}^{n} \mathbb{E}\left(r_{t}\right))^{12} - 1
\end{equation*}\\\\
where $r_{y}$ is the yearly return, $r_{m}$ is the monthly return, $r_{t}$ is the daily return, $n$ is trading days of that month, $\mathbb{E}(\cdot)$ is the expectation function.
\subsubsection{Portfolio return}
Given any set of risky assets and a set of weights that describe how the portfolio investment is
split, the general formula of expected return for $\mathrm{n}$ assets is:\\\\
\begin{equation*}
\mathbb{E}\left(r_{P}\right)=\sum_{i=1}^{n} w_{i} \mathbb{E}\left(r_{i}\right)
\end{equation*}\\\\
where $\sum_{i=1}^{n} w_{i}=1.0$; $n=$ the number of securities; $w_{i}$ is the proportion of the funds invested in security $i$; $r_{i}$ and $r_{P}$ are the return on ith security and portfolio $p$; $\mathbb{E}(\cdot)$ is the expectation function.\\\\
The return computation is nothing more than finding the weighted average return of the
securities included in the portfolio.\cite{chen2010portfolio}

\section{Result}
\subsection{Portfolio optimization}
 For the constraints of the quadratic programming optimization, the first constraint is that the sum of the weights equals 1, the second constraint is that the weight is greater than or equal to 0 (not allowed to sell short). The stock returns and standard deviation are calculated from the daily close price from January to September 2020. Considering different investment risk preferences, we found two optimal weights with the greatest Sharpe ratio and the smallest standard deviation. To calculate the Sharpe ratio, set the risk-free interest rate as 4\%.  \\\\
\begin{figure}[h]
\centering
\includegraphics[scale=0.7]{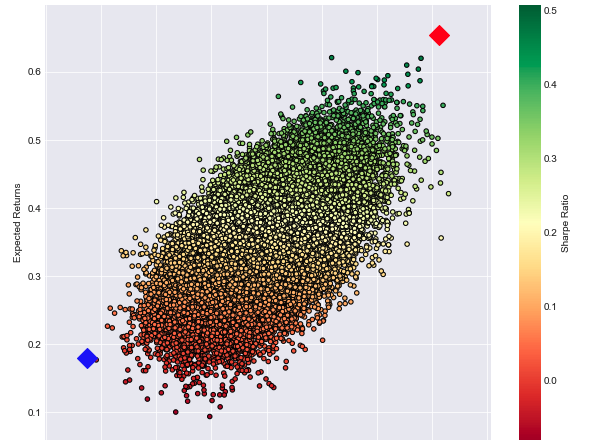}
\caption{Expected returns and Sharpe ratios for random portfolios.\\
\small
Red rhombus is the optimal portfolio with the greatest Sharpe ratio. Blue rhombus is the optimal portfolio with the smallest standard deviation.
\normalsize
}
\label{fig:markowitz}
\end{figure}\\\\
\noindent In Figure \ref{fig:markowitz}, the points are randomly produced 50,000 groups of weights for the 30 stocks. The optimal portfolio with the greatest Sharpe ratio has an expected yearly return of 65.44\% and a yearly volatility 0.27. The Sharpe ratio of this portfolio is 0.51. The optimal portfolio with the smallest standard deviation has an expected yearly return of 17.95\% and a yearly volatility 0.21. In the following research, we use the greatest Sharpe ratio portfolio for a high return. The Table \ref{table:monthly_return} shows the weights of each stock in this optimal portfolio.\\\\
As the result of optimization, the greatest weight is 0.0818 for 'SZ300122' and the smallest weight is 0.0049 for 'SZ000725'. The two stocks have monthly returns 14.98\% and -3.67\% in October respectively. The greater weights are given to 'SZ002475', 'SZ002594', 'SZ002607', 'SZ300122', 'SH600519', 'SH601888', 'SH601899', 'SH603288', in which the weights are over 0.05. In order of the stocks in the Table \ref{table:monthly_return} from top to bottom, they provide -0.02\%, 0.27\%, 0.12\%, -0.02\%, 0.45\%, 0.02\%, 0.61\%, -0.27\%, 2.75\%, 1.46\%, 0.86\%, 1.23\%, -0.04\%, -0.01\%, 0.84\%, 0.02\%, -0.09\%, -0.15\%, -0.01\%, -0.03\%, -0.05\%, -0.02\%, -0.05\%, 0, 0.01\%, 0.01\%, 0.76\%, -0.70\%, -0.02\%, 0.08\% respectively to the portfolio monthly return. Given the contributions to the portfolio, the top-performing stocks are 'SZ002594', 'SZ002607', 'SZ300015', among which contributions to the portfolio are over 1\% monthly. \\\\ 
By the optimal portfolio with the greatest Sharpe ratio, we have a monthly return of 8.02\% in October and a yearly return of 152.35\% for the origin holdings.  
\newpage 
\begin{table}[H]
    \centering
    \caption{30 stocks October daily return}
    \begin{tabular}{c|cccc}
\hline
 Ticker symbol & Weight & Close price on 2020-9-30 & Close price on 2020-10-30 & Monthly return (\%)\\
\hline
SZ000002 & 0.0110 & 28.02 & 27.49 & -1.89\\
SZ000333 & 0.0373 & 72.60 & 77.87 & 7.26\\
SZ000651 & 0.0128 & 53.30 & 58.43 & 9.62\\
SZ000725 & 0.0049 & 4.91  & 4.73  & -3.67\\
SZ000858 & 0.0425 & 221.00 & 244.35 & 10.57\\
SZ002352 & 0.0091 & 81.20 & 82.80 & 1.97\\
SZ002415 & 0.0345 & 38.11 & 44.90 & 17.82\\
SZ002475 & 0.0673 & 57.13 & 54.86 & -3.97 \\
SZ002594 & 0.0733 & 116.24 & 159.81 & 37.48\\
SZ002607 & 0.0692 & 32.63 & 39.52 & 21.12\\
SZ300015 & 0.0409 & 51.42 & 62.26 & 21.08\\
SZ300122 & 0.0818 & 139.31 & 160.18 & 14.98\\
SH600009 & 0.0090 & 68.78 & 66.10 & -3.90\\
SH600028 & 0.0058 & 3.91 & 3.90 & -2.56\\
SH600104 & 0.0348 & 19.13 & 23.15 & 24.01\\
SH600276 & 0.0226 & 89.82 & 88.84 & 1.09\\
SH600309 & 0.0068 & 69.30 & 78.49 & -13.26\\
SH600346 & 0.0368 & 18.56 & 19.30 & -3.99\\
SH600519 & 0.0782 & 1668.50 & 1670.02 & -0.09\\
SH601012 & 0.0200 & 75.01 & 75.99 & -1.31\\
SH601088 & 0.0435 & 16.47 & 16.65 & -1.09\\
SH601225 & 0.0047 & 8.39 & 8.75 & -4.29\\
SH601318 & 0.0259 & 76.26 & 77.83 & -2.06\\
SH601398 & 0.0140 & 4.92 & 4.92 &  0\\
SH601766 & 0.0060 & 5.49 & 5.39 & 1.82\\
SH601857 & 0.0075 & 4.11 & 4.07 &  0.97\\
SH601888 & 0.0719 & 222.94 & 199.40 & 10.56\\
SH601899 & 0.0513 & 6.15 & 6.99 & -13.66\\
SH601939 & 0.0094 & 6.15 & 6.29 & -2.28\\
SH603288 & 0.0672 & 162.10 & 160.20 &  1.17\\
\hline
    \end{tabular}
    \label{table:monthly_return}
\end{table}
\newpage
\subsection{Model}
To predict by MLP, we trained a new model for each trading day using the data before 11:20. For each model, lags of input time series to use as inputs are 60. The number of networks to train is 100, and the result is the ensemble forecast. To avoid the affection of outliers, the combination operator for forecasts is 'median.' The best number of hidden nodes is found by 5-fold cross-validation.\\\\
We use this prediction model to do paper trading. This model is used to predict the stock price of 13:00 to 14:50. With the price point at 14:50, we make trading decision by the difference between this point and the price at 11:20. If the prediction price is lower than the price at 11:20, we sell holding positions before the close in the morning and buy back at 14:50. If the prediction price is higher than the price at 11:20, we buy additional positions before the close in the morning and sell them at 14:50. Given the trading fee of 0.05\%, if the difference between the predicted price at 14:50 and the price at 11:20 is not more than 0.1\% of the latter, we will not trade on that day. To calculate the daily return, we use the close price on September 30th as origin position price. The October sum of daily return for each stock in the portfolio is as the Table \ref{table:daily_return}. The specific earnings and model prediction for each day of each stock are shown in Appendix. \\\\
In Table \ref{table:daily_return}, 'SZ000333', 'SZ000651', 'SZ000725', 'SZ000858', 'SZ002352', 'SZ002415', 'SZ002475', 'SZ002594', 'SZ002 607', 'SZ300015', 'SH600028', 'SH600309', 'SH600346', 'SH601088', 'SH601398', 'SH601857', 'SH603288' have positive returns. The number of positive returns stocks is 17 and the number of negative returns sto cks is 13. The top-performing stocks in the intraday trading strategy are 'SZ002594', 'SZ002607' and 'SZ002475' with the sum of daily return of 14.13\%, 10.10\%, and 7.09\% respectively. Particularly, they are given greater weights of 0.0733, 0.0692, and 0.0673.\\\\
For 'SZ002594', in the Table \ref{table:SZ002594} (See Appendix), the highest daily return is 6.27\% and the lowest return is -5.43\%. For 'SZ002607', in the Table \ref{table:SZ002607} (See Appendix), the highest daily return is 3.88\% and the lowest return is -2.21\%. For 'SZ002478', in the Table \ref{table:SZ002475} (See Appendix), the highest daily return is 2.94\% and the lowest return is -0.68\%. All three stocks have posted strong gains throughout the month, but have also posted big losses in some trading sessions.\\\\
Stocks with high intraday volatility, such as 'SZ300122', 'SH600104' and 'SH601318' are prone to large losses. The October sum of daily returns for the three stocks are -11.48\%, -7.11\% and -4.04\%, respectively. The weights for these three stocks are 0.0818, 0.0348, 0.0259.\\\\
For 'SZ300122', in the Table \ref{table:SZ300122} (See Appendix), the highest daily return is 8.48\% and the lowest return is -7.90\%. For 'SH600104', in the Table \ref{table:SH600104} (See Appendix), the highest daily return is 1.05\% and the lowest return is -3.08\%. For 'SH601318', in the Table \ref{table:SH601318} (See Appendix), the highest daily return is 1.08\% and the lowest return is -1.32\%. All three stocks have posted strong gains throughout the month, but have also posted big losses in some trading sessions.\\\\
In the prediction, the model is able to predict accurately when price rises or falls periodically. But in periods of sustained rise or fall, when models typically predict reversion to the mean, it tends to generate large trading losses.\\\\
Combining the prediction model with the intraday trading strategy, we have a return of 1.41\% in October and a yearly return of 18.24\%. 
\newpage
\begin{table}[H]
    \centering
    \caption{30 stocks October daily return}
    \begin{tabular}{cc}
\hline
 Ticker symbol& October sum of daily return (\%) \\
\hline
SZ000002 & -1.96\\
SZ000333 & 3.66 \\
SZ000651 & 2.35 \\
SZ000725 & 1.22\\
SZ000858 & 0.88\\
SZ002352 & 2.38\\
SZ002415 & 0.25\\
SZ002475 & 7.09 \\
SZ002594 & 14.13\\
SZ002607 & 10.10\\
SZ300015 & 2.40\\
SZ300122 & -11.48\\
SH600009 & -1.45\\
SH600028 & 1.29\\
SH600104 & -7.11\\
SH600276 & -1.51\\
SH600309 & 5.25\\
SH600346 & 6.74\\
SH600519 & -3.41\\
SH601012 & -1.15\\
SH601088 & 0.14\\
SH601225 & -0.7\\
SH601318 & -4.04\\
SH601398 & 0.61\\
SH601766 & -0.19\\
SH601857 & 1.69\\
SH601888 & -0.72\\
SH601899 & -1.12 \\
SH601939 & -1.78 \\
SH603288 & 4.72\\
\hline
    \end{tabular}
    \label{table:daily_return}
\end{table}
\newpage
\subsection{Portfolio return}
Total portfolio returns are divided into origin position gains and intraday trading gains. As mentioned before, the origin position gains for October is 8.02\% and the intraday trading gains for October is 1.41\%. The total portfolio returns for October is 9.43\% and the yearly return is 194.87\%.

\section{Discussion}
The current study's purpose was to combine the prediction model and practical intraday trading strategy under the T+1 constraint in China stock market. This study has shown that the prediction model and strategy is profitable for trading. The findings will be of interest to the Fund company investing in China stock market and interested in the intraday trading under T+1.\\\\
More broadly, research is also needed to determine some practical trading problems. The first problem is to set a stop loss. When we find the prediction trend is reverse to the actual trend and the loss is too large in a day, like 3\%, we should trade ahead of time to avoid a larger loss. The second problem is to set a higher threshold for the decision of trading. Because the trading fee is 0.05\% per time, it will cause loss if we trade too many times, even the strategy is profitable. In this article, we set the threshold as 0.1\%. If the difference between the prediction price at 14:50 and the real price at 11:20 is not more than or less than 0.1\% of the origin position price, we will not trade on that day. To reduce the trading fee, we can increase the threshold to 0.2\% or more.


\bibliography{bibliography.bib}

\newpage
\appendix
\section*{Appendix}
\begin{table}[H]
    \centering
        \caption{Ticker symbol reference}
    \begin{tabular}{c|cc}
\hline
Number of Stock & Ticker symbol& Company\\
\hline
01 & SZ000002 & CHINA VANKE CO., Ltd.\\
02 & SZ000333 & Midea Group \\
03 & SZ000651 & Gree Electric Appliances Inc. of Zhuhai \\
04 & SZ000725 & Boe Technology Group Co., Ltd.\\
05 & SZ000858 & Wuliangye Yibin Co.,Ltd. \\
06 & SZ002352 & SF Express (Group) Co., Ltd.\\
07 & SZ002415 & Hangzhou Hikvision Digital Technology Co., Ltd.\\
08 & SZ002475 & Shenzhen Luxshare Precision Industry Co.,Ltd. \\
09 & SZ002594 & BYD Co., Ltd.\\
10 & SZ002607 & Offcn Education Technology Co., Ltd.\\
11 & SZ300015 & Aier Eye Hospital Group Co., Ltd.\\
12 & SZ300122 & Chongqing Zhifei Biological Products Co., Ltd.\\
13 & SH600009 & Shanghai Airport Authority\\
14 & SH600028 & Sinopec, the China Petroleum and Chemical Corporation\\
15 & SH600104 & SAIC Motor Corporation Limited\\
16 & SH600276 & Jiangsu Hengrui Medicine Co., Ltd.\\
17 & SH600309 & Wanhua Chemical Group Co., Ltd.\\
18 & SH600346 & Hengli Group\\
19 & SH600519 & Kweichow Moutai Co., Ltd.\\
20 & SH601012 & LONGi Green Energy Technology Co.,Ltd.\\
21 & SH601088 & China Shenhua Energy Co., Ltd.\\
22 & SH601225 & Shaanxi Coal and Chemical Industry Group Co., Ltd.\\
23 & SH601318 & Ping An Insurance (Group) Company of China, Ltd.\\
24 & SH601398 & Industrial and Commercial Bank of China Limited\\
25 & SH601766 & CRRC Corporation Limited\\
26 & SH601857 & China National Petroleum Corporation\\
27 & SH601888 & China Tourism Group Duty Free Corporation Limited.\\
28 & SH601899 & Zijin Mining Group Co., Limited \\
29 & SH601939 & China Construction Bank Corporation \\
30 & SH603288 & Foshan Haitian Flavouring \& Food Co. Ltd\\
\hline
    \end{tabular}
    \label{table:ticker}
\end{table}

\begin{figure}[H]
\centering
\includegraphics[scale=0.2]{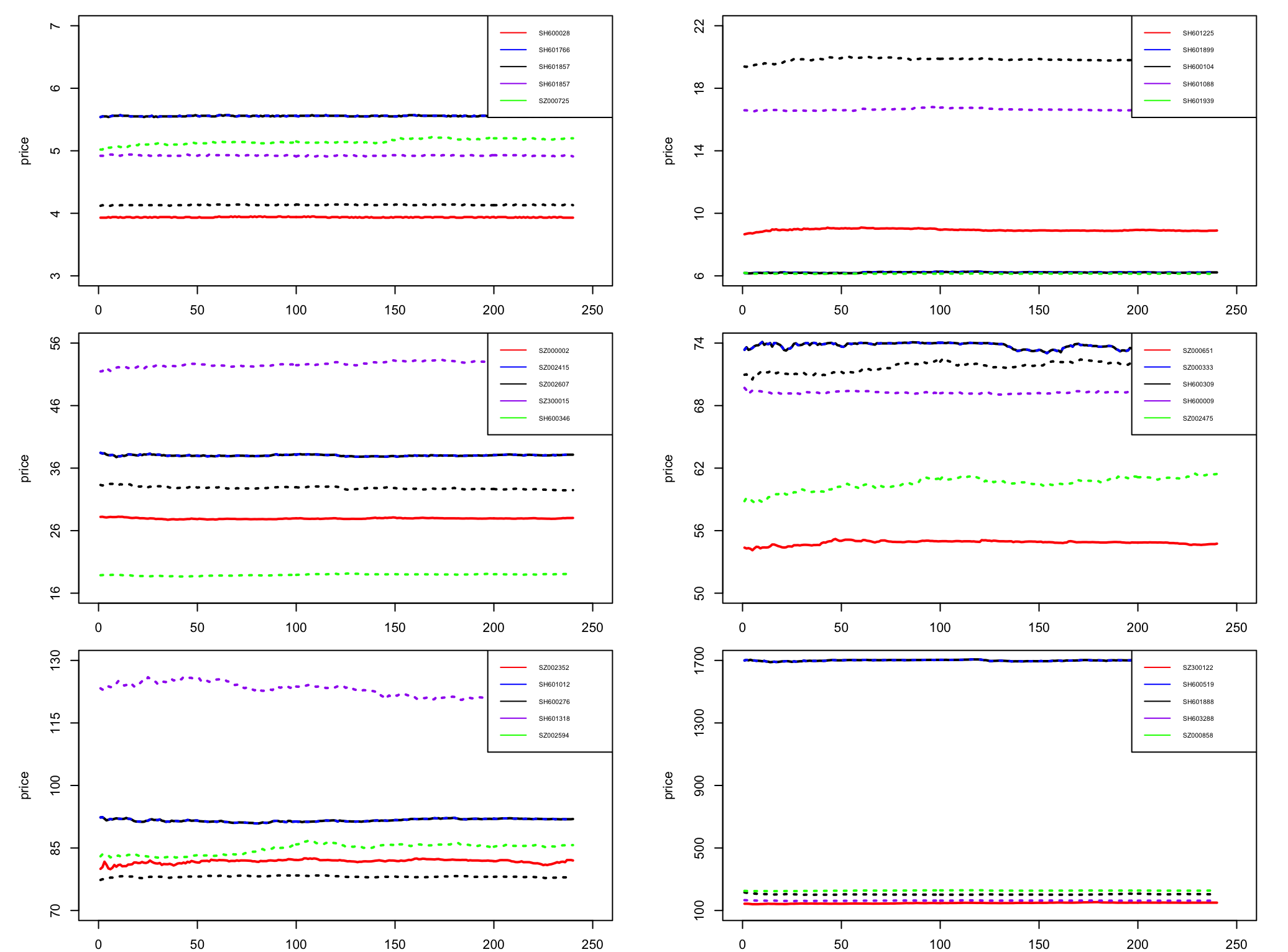}
\caption{Minute close price on 2020-10-09}
\label{fig:minute_data}
\end{figure}

\begin{figure}[H]
\centering
\includegraphics[scale=0.2]{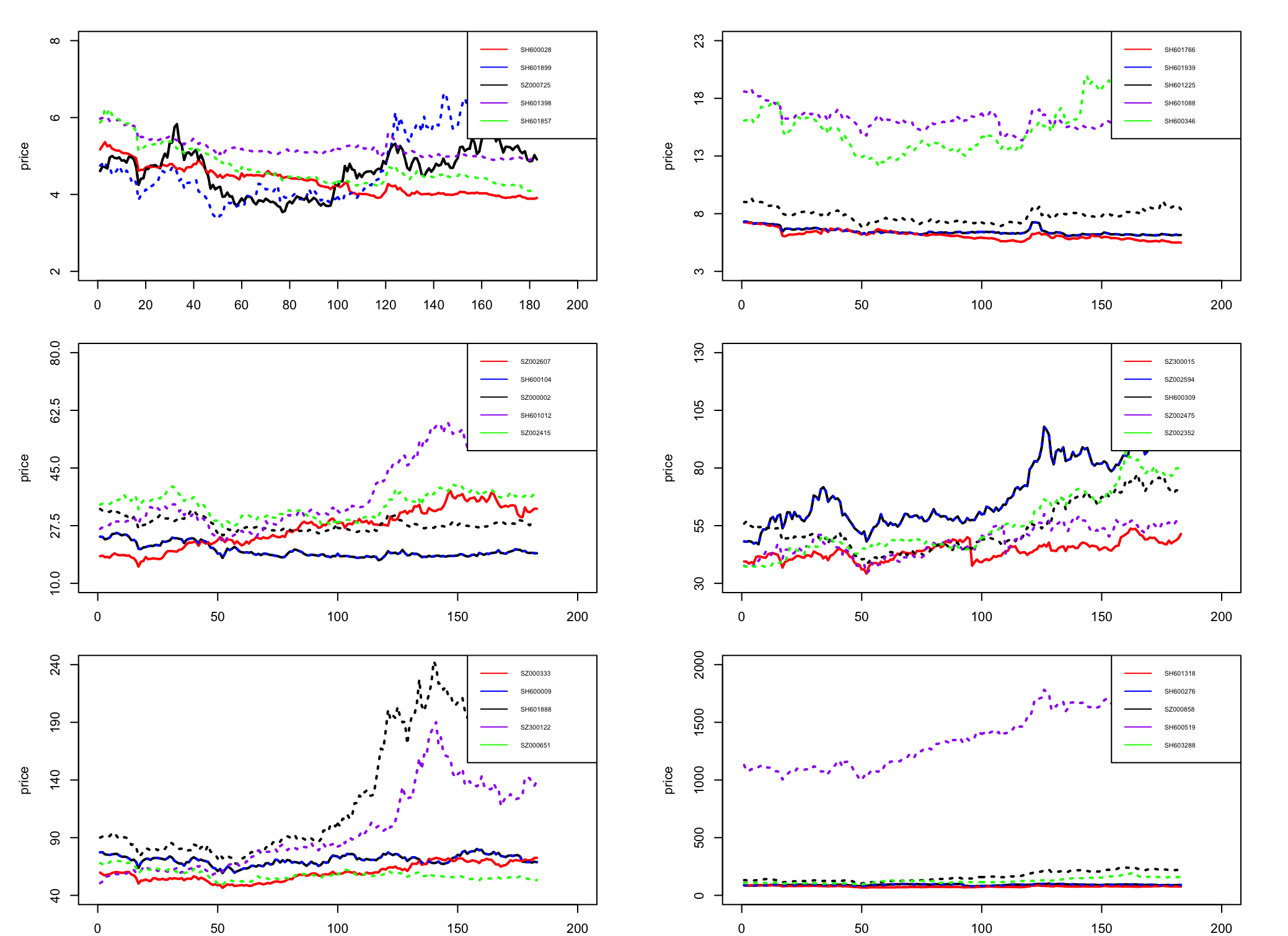}
\caption{Daily close price from January to September 2020}
\label{fig:daily_data}
\end{figure}

\begin{table}
    \centering
        \caption{SZ000002 daily return}
    \begin{tabular}{c|cccc}
\hline
Date & Price at 11:20 & Real price at 14:50 & Predicted price at 14:50 & Daily return (\%)\\
\hline
2020-10-9  & 27.90 & 27.90 & 27.89 & 0\\
2020-10-12 & 28.37 & 28.26 & 28.39 & 0\\
2020-10-13 & 27.90 & 27.95 & 27.75 & -0.18 \\
2020-10-14 & 27.75 & 27.79 & 27.49 & -0.14\\
2020-10-15 & 27.77 & 27.73 & 27.80 & -0.14\\
2020-10-16 & 27.81 & 27.83 & 28.20 & 0.07\\
2020-10-19 & 27.96 & 27.74 & 28.11 & -0.79\\
2020-10-20 & 27.42 & 27.45 & 27.44 & 0\\
2020-10-21 & 27.35 & 27.60 & 27.22 & -0.89\\
2020-10-22 & 27.79 & 27.99 & 27.96 & 0.71\\
2020-10-23 & 27.93 & 27.95 & 27.97 & 0.07\\
2020-10-26 & 27.99 & 27.99 & 27.96 & 0\\
2020-10-27 & 27.61 & 27.50 & 27.69 & -0.39\\
2020-10-28 & 27.00 & 26.98 & 26.98 & 0\\
2020-10-29 & 27.09 & 27.45 & 27.24 & 1.28\\
2020-10-30 & 27.88 & 27.44 & 28.36 & -1.57\\
\hline
    \end{tabular}
    \label{table:SZ000002}
\end{table}

\begin{table}
    \centering
    \caption{SZ000333 daily return}
    \begin{tabular}{c|cccc}
\hline
Date & Price at 11:20 & Real price at 14:50 & Predicted price at 14:50 & Daily return (\%)\\
\hline
2020-10-9  & 73.88 & 73.38 & 73.96 & -0.69\\
2020-10-12 & 73.16 & 73.70 & 73.15 & 0\\
2020-10-13 & 74.90 & 75.91 & 74.92 & 0 \\
2020-10-14 & 75.27 & 75.41 & 75.32 & 0\\
2020-10-15 & 76.73 & 76.29 & 76.54 & 0.6\\
2020-10-16 & 74.73 & 75.10 & 75.02 & 0.51\\
2020-10-19 & 75.12 & 74.84 & 74.41 & 0.39\\
2020-10-20 & 76.03 & 76.19 & 75.82 & -0.22\\
2020-10-21 & 77.45 & 77.31 & 79.25 & -0.19\\
2020-10-22 & 77.46 & 78.22 & 77.35 & -1.05\\
2020-10-23 & 77.36 & 75.64 & 77.44 & 0\\
2020-10-26 & 76.37 & 76.53 & 76.91 & 0.22\\
2020-10-27 & 75.95 & 75.99 & 75.91 & 0\\
2020-10-28 & 76.48 & 76.89 & 74.73 & -0.56\\
2020-10-29 & 79.78 & 81.64 & 82.51 & 2.56\\
2020-10-30 & 79.28 & 77.76 & 77.31 & 2.09\\
\hline
    \end{tabular}
    \label{table:SZ000333}
\end{table}

\begin{table}
    \centering
    \caption{SZ000651 daily return}
    \begin{tabular}{c|cccc}
\hline
Date & Price at 11:20 & Real price at 14:50 & Predicted price at 14:50 & Daily return (\%)\\
\hline
2020-10-9  & 54.99 & 54.65 & 55.02 & 0\\
2020-10-12 & 55.77 & 55.77 & 55.70 & 0\\
2020-10-13 & 55.76 & 55.89 & 55.66 & -0.24 \\
2020-10-14 & 57.54 & 57.64 & 57.67 & 0.19\\
2020-10-15 & 58.37 & 57.91 & 58.37 & 0\\
2020-10-16 & 58.21 & 57.73 & 58.33 & -0.90\\
2020-10-19 & 57.81 & 57.25 & 57.63 & 1.05\\
2020-10-20 & 57.56 & 57.64 & 57.72 & 0.15\\
2020-10-21 & 57.87 & 57.92 & 57.88 & 0\\
2020-10-22 & 58.35 & 58.56 & 58.33 & 0\\
2020-10-23 & 58.28 & 58.51 & 57.61 & -0.43\\
2020-10-26 & 58.00 & 58.10 & 57.28 & -0.19\\
2020-10-27 & 57.63 & 57.34 & 57.73 & -0.54\\
2020-10-28 & 57.26 & 57.31 & 57.39 & 0.09\\
2020-10-29 & 57.28 & 58.12 & 57.92 & 1.58\\
2020-10-30 & 59.30 & 58.45 & 59.22 & 1.59\\
\hline
    \end{tabular}
    
    \label{table:SZ000651}
\end{table}

\begin{table}
    \centering
    \caption{SZ000725 daily return}
    \begin{tabular}{c|cccc}
\hline
Date & Price at 11:20 & Real price at 14:50 & Predicted price at 14:50 & Daily return (\%)\\
\hline
2020-10-9  & 5.13 & 5.18 & 5.27 & 1.02\\
2020-10-12 & 5.24 & 5.23 & 5.23 & 0.20\\
2020-10-13 & 5.16 & 5.17 & 5.15 & -0.20 \\
2020-10-14 & 5.06 & 5.05 & 5.12 & -0.20\\
2020-10-15 & 5.00 & 4.97 & 5.01 & -0.61\\
2020-10-16 & 4.89 & 4.91 & 4.88 & -0.41\\
2020-10-19 & 4.88 & 4.85 & 4.88 & 0\\
2020-10-20 & 4.79 & 4.92 & 4.80 & 2.65\\
2020-10-21 & 4.80 & 4.85 & 4.81 & 1.02\\
2020-10-22 & 4.89 & 4.88 & 4.90 & -0.20\\
2020-10-23 & 4.83 & 4.79 & 4.83 & 0\\
2020-10-26 & 4.79 & 4.79 & 4.80 & 0\\
2020-10-27 & 4.71 & 4.74 & 4.70 & -0.61\\
2020-10-28 & 4.72 & 4.79 & 4.70 & -1.43\\
2020-10-29 & 4.87 & 4.83 & 5.13 & -0.81\\
2020-10-30 & 4.87 & 4.83 & 5.13 & -0.81\\
\hline
    \end{tabular}
    
    \label{table:SZ000725}
\end{table}

\begin{table}
    \centering
    \caption{SZ000858 daily return}
    \begin{tabular}{c|cccc}
\hline
Date & Price at 11:20 & Real price at 14:50 & Predicted price at 14:50 & Daily return (\%)\\
\hline
2020-10-9  & 229.23 & 226.94 & 234.61 & -1.04\\
2020-10-12 & 237.99 & 241.99 & 245.37 & 1.81\\
2020-10-13 & 241.61 & 242.52 & 241.61 & 0 \\
2020-10-14 & 239.39 & 239.61 & 228.42 & -0.09\\
2020-10-15 & 238.81 & 238.85 & 238.82 & 0\\
2020-10-16 & 237.12 & 236.93 & 238.11 & -0.08\\
2020-10-19 & 236.27 & 235.69 & 237.12 & -0.26\\
2020-10-20 & 239.20 & 240.98 & 239.53 & 0.81\\
2020-10-21 & 240.94 & 241.86 & 241.02 & 0\\
2020-10-22 & 240.30 & 242.61 & 240.70 & 1.05\\
2020-10-23 & 240.66 & 237.10 & 241.41 & -1.61\\
2020-10-26 & 233.87 & 233.17 & 235.84 & -0.32\\
2020-10-27 & 232.54 & 234.10 & 232.51 & 0\\
2020-10-28 & 239.09 & 240.40 & 243.81 & 0.59\\
2020-10-29 & 253.13 & 251.03 & 266.74 & -0.95\\
2020-10-30 & 254.94 & 243.75 & 245351 & 0.99\\
\hline
    \end{tabular}
    
    \label{table:SZ000858}
\end{table}

\begin{table}
    \centering
    \caption{SZ002352 daily return}
    \begin{tabular}{c|cccc}
\hline
Date & Price at 11:20 & Real price at 14:50 & Predicted price at 14:50 & Daily return (\%)\\
\hline
2020-10-9  & 86.20 & 85.37 & 87.05 & -1.02\\
2020-10-12 & 86.40 & 86.91 & 85.96 & -0.63\\
2020-10-13 & 89.97 & 92.26 & 89.54 & -2.82 \\
2020-10-14 & 92.07 & 92.82 & 92.44 & 0.92\\
2020-10-15 & 89.75 & 90.3 & 88.93 & -0.68\\
2020-10-16 & 89.30 & 89.66 & 89.43 & 0.44\\
2020-10-19 & 89.59 & 89.37 & 89.98 & -0.27\\
2020-10-20 & 88.48 & 90.60 & 88.59 & 2.61\\
2020-10-21 & 89.44 & 89.50 & 89.33 & -0.07\\
2020-10-22 & 86.37 & 88.01 & 86.06 & -2.02\\
2020-10-23 & 86.49 & 85.02 & 84.93 & 1.81\\
2020-10-26 & 87.98 & 87.17 & 87.82 & 1.00\\
2020-10-27 & 86.14 & 87.04 & 86.17 & 0\\
2020-10-28 & 86.18 & 86.82 & 85.95 & -0.79\\
2020-10-29 & 85.54 & 86.33 & 85.78 & 0.97\\
2020-10-30 & 84.88 & 82.50 & 83.42 & 2.93\\
\hline
    \end{tabular}
    
    \label{table:SZ002352}
\end{table}

\begin{table}
    \centering
    \caption{SZ002415 daily return}
    \begin{tabular}{c|cccc}
\hline
Date & Price at 11:20 & Real price at 14:50 & Predicted price at 14:50 & Daily return (\%)\\
\hline
2020-10-9  & 38.19 & 38.10 & 38.01 & 0.24\\
2020-10-12 & 37.96 & 38.65 & 37.96 & 0\\
2020-10-13 & 39.01 & 39.03 & 38.75 & -0.05 \\
2020-10-14 & 38.80 & 38.97 & 39.04 & 0.45\\
2020-10-15 & 39.63 & 39.43 & 39.55 & 0.52\\
2020-10-16 & 38.45 & 38.52 & 38.67 & 0.18\\
2020-10-19 & 38.98 & 39.26 & 39.04 & 0.73\\
2020-10-20 & 39.18 & 39.19 & 39.15 & 0\\
2020-10-21 & 39.39 & 39.20 & 39.32 & 0.5\\
2020-10-22 & 39.39 & 39.24 & 39.03 & 0.39\\
2020-10-23 & 40.24 & 39.00 & 40.34 & -3.25\\
2020-10-26 & 43.00 & 42.56 & 46.38 & -1.15\\
2020-10-27 & 44.00 & 44.09 & 45.94 & 0.24\\
2020-10-28 & 44.31 & 44.22 & 44.15 & 0.24\\
2020-10-29 & 44.55 & 45.01 & 44.72 & 1.21\\
2020-10-30 & 46.00 & 45.01 & 45.99 & 0\\
\hline
    \end{tabular}
    
    \label{table:SZ002415}
\end{table}

\begin{table}
    \centering
    \caption{SZ002475 daily return}
    \begin{tabular}{c|cccc}
\hline
Date & Price at 11:20 & Real price at 14:50 & Predicted price at 14:50 & Daily return (\%)\\
\hline
2020-10-9  & 61.13 & 61.36 & 61.49 & 0.40\\
2020-10-12 & 61.04 & 61.98 & 61.21 & 1.65\\
2020-10-13 & 61.75 & 61.34 & 61.08 & 0.72 \\
2020-10-14 & 59.44 & 59.88 & 59.80 & 0.77\\
2020-10-15 & 60.25 & 59.92 & 60.01 & 0.58\\
2020-10-16 & 58.11 & 58.40 & 56.08 & -0.51\\
2020-10-19 & 58.91 & 58.49 & 58.96 & 0\\
2020-10-20 & 58.75 & 59.17 & 58.73 & 0\\
2020-10-21 & 57.62 & 57.74 & 57.62 & 0\\
2020-10-22 & 57.15 & 57.42 & 57.28 & 0.47\\
2020-10-23 & 57.20 & 56.78 & 57.40 & -0.74\\
2020-10-26 & 57.58 & 57.21 & 57.67 & -0.65\\
2020-10-27 & 56.56 & 57.50 & 56.70 & 1.65\\
2020-10-28 & 59.10 & 59.49 & 58.63 & -0.68\\
2020-10-29 & 58.71 & 58.43 & 58.42 & 0.49\\
2020-10-30 & 56.64 & 54.96 & 55.31 & 2.94\\
\hline
    \end{tabular}
    
    \label{table:SZ002475}
\end{table}
\begin{table}
    \centering
    \caption{SZ002594 daily return}
    \begin{tabular}{c|cccc}
\hline
Date & Price at 11:20 & Real price at 14:50 & Predicted price at 14:50 & Daily return (\%)\\
\hline
2020-10-9  & 123.68 & 120.00 & 123.56 & 0\\
2020-10-12 & 121.03 & 128.36 & 121.39 & 6.31\\
2020-10-13 & 129.20 & 129.29 & 128.69 & -0.08 \\
2020-10-14 & 130.32 & 130.89 & 132.67 & 0.49\\
2020-10-15 & 134.38 & 131.94 & 134.27 & 0\\
2020-10-16 & 125.88 & 127.69 & 119.85 & -1.56\\
2020-10-19 & 127.25 & 127.10 & 127.43 & -0.13\\
2020-10-20 & 131.51 & 138.80 & 132.12 & 6.27\\
2020-10-21 & 137.09 & 138.00 & 136.99 & 0\\
2020-10-22 & 140.50 & 144.52 & 144.37 & 3.46\\
2020-10-23 & 141.00 & 137.52 & 142.19 & -2.99\\
2020-10-26 & 141.86 & 139.95 & 141.92 & 0\\
2020-10-27 & 137.85 & 140.37 & 137.88 & 0\\
2020-10-28 & 147.91 & 151.40 & 154.99 & 3\\
2020-10-29 & 158.02 & 163.59 & 159.63 & 4.79\\
2020-10-30 & 166.48 & 160.17 & 167.42 & -5.43\\
\hline
    \end{tabular}
    
    \label{table:SZ002594}
\end{table}

\begin{table}
    \centering
    \caption{SZ002607 daily return}
    
    \begin{tabular}{c|cccc}
\hline
Date & Price at 11:20 & Real price at 14:50 & Predicted price at 14:50 & Daily return (\%)\\
\hline
2020-10-9  & 33.07 & 32.51 & 32.95 & 1.72\\
2020-10-12 & 31.94 & 32.07 & 31.02 & -0.40\\
2020-10-13 & 32.41 & 32.94 & 32.87 & 1.62 \\
2020-10-14 & 36.25 & 36.25 & 36.19 & 0\\
2020-10-15 & 36.32 & 36.69 & 35.98 & -1.13\\
2020-10-16 & 37.88 & 38.64 & 37.95 & 2.33\\
2020-10-19 & 37.70 & 38.17 & 37.76 & 1.44\\
2020-10-20 & 37.84 & 37.73 & 37.98 & -0.34\\
2020-10-21 & 38.08 & 38.09 & 37.99 & -0.03\\
2020-10-22 & 38.67 & 38.71 & 38.76 & 0.12\\
2020-10-23 & 39.31 & 37.89 & 39.27 & 4.35\\
2020-10-26 & 37.23 & 37.34 & 36.99 & -0.34\\
2020-10-27 & 38.10 & 38.06 & 38.17 & -0.12\\
2020-10-28 & 39.47 & 38.75 & 40.80 & -2.21\\
2020-10-29 & 39.87 & 39.62 & 41.18 & -0.77\\
2020-10-30 & 40.58 & 39.32 & 40.52 & 3.86\\
\hline
    \end{tabular}
    \label{table:SZ002607}
    
\end{table}

\begin{table}
    \centering
    \caption{SZ300015 daily return}
    \begin{tabular}{c|cccc}
\hline
Date & Price at 11:20 & Real price at 14:50 & Predicted price at 14:50 & Daily return (\%)\\
\hline
2020-10-9  & 52.54 & 53.28 & 52.53 & 0\\
2020-10-12 & 55.25 & 57.47 & 57.36 & 4.32\\
2020-10-13 & 56.80 & 58.12 & 54.84 & -2.57 \\
2020-10-14 & 57.90 & 57.72 & 57.49 & 0.35\\
2020-10-15 & 57.15 & 57.16 & 56.24 & -0.02\\
2020-10-16 & 56.55 & 57.18 & 56.60 & 0\\
2020-10-19 & 56.40 & 56.20 & 56.78 & -0.39\\
2020-10-20 & 56.91 & 57.43 & 56.99 & 1.01\\
2020-10-21 & 57.67 & 58.25 & 57.44 & -1.13\\
2020-10-22 & 57.78 & 58.80 & 57.76 & 0\\
2020-10-23 & 58.58 & 56.96 & 58.65 & -3.15\\
2020-10-26 & 56.91 & 57.00 & 56.54 & -0.18\\
2020-10-27 & 60.30 & 61.50 & 60.95 & 2.33\\
2020-10-28 & 62.70 & 63.14 & 64.55 & 0.86\\
2020-10-29 & 62.75 & 62.58 & 62.57 & 0.33\\
2020-10-30 & 62.38 & 62.05 & 62.12 & 0.64\\
\hline
    \end{tabular}
    
    \label{table:SZ300015}
\end{table}

\begin{table}
    \centering
    \caption{SZ300122 daily return}
    \begin{tabular}{c|cccc}
\hline
Date & Price at 11:20 & Real price at 14:50 & Predicted price at 14:50 & Daily return (\%)\\
\hline
2020-10-9  & 148.41 & 149.98 & 150.60 & 1.13\\
2020-10-12 & 153.97 & 156.10 & 150.39 & -1.53\\
2020-10-13 & 163.38 & 156.41 & 167.50 & -5.00 \\
2020-10-14 & 160.90 & 157.95 & 161.15 & -2.12\\
2020-10-15 & 154.10 & 156.55 & 145.62 & -1.76\\
2020-10-16 & 157.74 & 161.87 & 159.79 & 2.96\\
2020-10-19 & 159.57 & 158.60 & 159.76 & -0.70\\
2020-10-20 & 155.85 & 160.01 & 155.50 & -2.99\\
2020-10-21 & 164.12 & 162.61 & 164.03 & 0\\
2020-10-22 & 154.45 & 153.79 & 155.32 & -0.47\\
2020-10-23 & 153.80 & 141.99 & 153.26 & 8.48\\
2020-10-26 & 143.25 & 142.92 & 146.33 & -0.24\\
2020-10-27 & 143.78 & 146.30 & 143.25 & -1.81\\
2020-10-28 & 145.28 & 147.06 & 144.63 & -1.28\\
2020-10-29 & 151.00 & 162.00 & 147.61 & -7.90\\
2020-10-30 & 160.74 & 158.30 & 160.07 & 1.75\\
\hline
    \end{tabular}
    
    \label{table:SZ300122}
\end{table}

\begin{table}
    \centering
    \caption{SH600009 daily return}
    \begin{tabular}{c|cccc}
\hline
Date & Price at 11:20 & Real price at 14:50 & Predicted price at 14:50 & Daily return (\%)\\
\hline
2020-10-9  & 69.25 & 69.27 & 68.98 & -0.03\\
2020-10-12 & 69.89 & 69.91 & 69.81 & -0.03\\
2020-10-13 & 69.73 & 69.92 & 68.98 & -0.28 \\
2020-10-14 & 69.19 & 70.38 & 69.17 & 0\\
2020-10-15 & 70.00 & 69.62 & 69.55 & 0.55\\
2020-10-16 & 69.55 & 69.50 & 69.75 & -0.07\\
2020-10-19 & 69.32 & 69.00 & 69.50 & -0.47\\
2020-10-20 & 69.44 & 69.52 & 70.21 & 0.12\\
2020-10-21 & 69.36 & 69.69 & 69.47 & 0.48\\
2020-10-22 & 68.29 & 67.84 & 68.23 & 0\\
2020-10-23 & 67.31 & 67.16 & 67.49 & -0.22\\
2020-10-26 & 67.20 & 67.38 & 67.18 & 0\\
2020-10-27 & 66.31 & 66.23 & 66.52 & -0.12\\
2020-10-28 & 66.58 & 67.09 & 66.53 & 0\\
2020-10-29 & 66.31 & 66.3 & 66.77 & -0.01\\
2020-10-30 & 67.08 & 66.14 & 67.62 & -1.37\\
\hline
    \end{tabular}
    
    \label{table:SH600009}
\end{table}

\begin{table}
    \centering
    \caption{SH600028 daily return}
    \begin{tabular}{c|cccc}
\hline
Date & Price at 11:20 & Real price at 14:50 & Predicted price at 14:50 & Daily return (\%)\\
\hline
2020-10-9  & 3.94 & 3.94 & 3.94 & 0\\
2020-10-12 & 3.95 & 3.96 & 4.03 & 0.26\\
2020-10-13 & 3.93 & 3.94 & 3.89 & -0.26\\
2020-10-14 & 3.92 & 3.90 & 3.91 & 0.51\\
2020-10-15 & 3.92 & 3.91 & 3.92 & -0.26\\
2020-10-16 & 3.94 & 3.93 & 3.93 & 0.26\\
2020-10-19 & 3.92 & 3.92 & 3.92 & 0\\
2020-10-20 & 3.92 & 3.92 & 3.92 & 0\\
2020-10-21 & 3.92 & 3.93 & 3.99 & 0.26\\
2020-10-22 & 3.93 & 3.92 & 3.92 & 0.26\\
2020-10-23 & 3.90 & 3.90 & 3.93 & 0\\
2020-10-26 & 3.90 & 3.90 & 3.90 & 0\\
2020-10-27 & 3.88 & 3.88 & 3.88 & 0\\
2020-10-28 & 3.86 & 3.87 & 3.87 & 0.26\\
2020-10-29 & 3.88 & 3.88 & 3.87 & 0\\
2020-10-30 & 3.92 & 3.9 & 3.92 & 0\\
\hline
    \end{tabular}
    
    \label{table:SH600028}
\end{table}

\begin{table}
    \centering
    \caption{SH600104 daily return}
    \begin{tabular}{c|cccc}
\hline
Date & Price at 11:20 & Real price at 14:50 & Predicted price at 14:50 & Daily return (\%)\\
\hline
2020-10-9  & 19.87 & 19.77 & 19.29 & 0.52\\
2020-10-12 & 19.96 & 20.07 & 20.00 & 0.58\\
2020-10-13 & 20.05 & 20.65 & 20.02 & -3.14 \\
2020-10-14 & 20.34 & 20.53 & 20.33 & 0\\
2020-10-15 & 20.71 & 20.51 & 20.83 & -1.05\\
2020-10-16 & 20.39 & 20.59 & 20.43 & 1.05\\
2020-10-19 & 20.39 & 20.37 & 20.51 & -0.10\\
2020-10-20 & 20.80 & 21.11 & 20.81 & 0\\
2020-10-21 & 21.41 & 21.57 & 21.39 & -0.84\\
2020-10-22 & 21.17 & 21.19 & 21.66 & 0.10\\
2020-10-23 & 21.49 & 21.05 & 21.49 & 0\\
2020-10-26 & 21.26 & 21.18 & 21.36 & -0.42\\
2020-10-27 & 21.13 & 21.12 & 21.20 & -0.05\\
2020-10-28 & 21.59 & 21.67 & 21.91 & 0.42\\
2020-10-29 & 22.61 & 22.82 & 22.52 & -1.10\\
2020-10-30 & 23.65 & 23.06 & 23.79 & -3.08\\
\hline
    \end{tabular}
    
    \label{table:SH600104}
\end{table}

\begin{table}
    \centering
    \caption{SH600276 daily return}
    \begin{tabular}{c|cccc}
\hline
Date & Price at 11:20 & Real price at 14:50 & Predicted price at 14:50 & Daily return (\%)\\
\hline
2020-10-9  & 91.31  &  91.97  &  91.33  &  0\\
2020-10-12 & 93.85  &  94.03  &  94.59  &  0.20\\
2020-10-13 & 93.66  &  93.75  &  93.46  &  -0.10\\
2020-10-14 & 93.24  &  92.95  &  93.27  &  0\\
2020-10-15 & 93.23  &  93.34  &  92.97  &  -0.12\\
2020-10-16 & 92.56  &  92.83  &  91.00  &  -0.30\\
2020-10-19 & 91.83  &  91.27  &  91.99  &  -0.62\\
2020-10-20 & 90.51  &  91.13  &  90.22  &  -0.69\\
2020-10-21 & 89.64  &  89.50  &  89.64  &  0\\
2020-10-22 & 89.31  &  89.39  &  89.15  &  -0.09\\
2020-10-23 & 89.20  &  88.29  &  89.27  &  0\\
2020-10-26 & 89.01  &  89.20  &  89.03  &  0\\
2020-10-27 & 88.96  &  89.23  &  89.21  &  0.30\\
2020-10-28 & 90.84  &  91.30  &  90.94  &  0.51\\
2020-10-29 & 91.05  &  91.50  &  90.77  &  -0.50\\
2020-10-30 & 88.56  &  88.65  &  87.35  &  -0.10\\
\hline
    \end{tabular}
    
    \label{table:SH600276}
\end{table}

\begin{table}
    \centering
    \caption{SH600309 daily return}
    \begin{tabular}{c|cccc}
\hline
Date & Price at 11:20 & Real price at 14:50 & Predicted price at 14:50 & Daily return (\%)\\
\hline
2020-10-9  & 72.02  &  72.16  &  72.16  &  0.20\\
2020-10-12 & 76.01  &  77.03  &  78.66  &  1.47\\
2020-10-13 & 77.02  &  79.37  &  76.97  &  0\\
2020-10-14 & 80.48  &  80.24  &  80.63  &  -0.35\\
2020-10-15 & 79.70  &  79.61  &  79.65  &  0\\
2020-10-16 & 78.75  &  79.73  &  79.40  &  1.41\\
2020-10-19 & 79.02  &  79.30  &  79.36  &  0.40\\
2020-10-20 & 79.11  &  80.01  &  79.31  &  1.30\\
2020-10-21 & 79.51  &  79.59  &  79.46  &  0\\
2020-10-22 & 78.42  &  78.91  &  78.54  &  0.71\\
2020-10-23 & 79.73  &  77.40  &  81.01  &  -3.36\\
2020-10-26 & 76.34  &  76.32  &  77.01  &  -0.03\\
2020-10-27 & 77.33  &  77.60  &  77.22  &  -0.39\\
2020-10-28 & 79.78  &  82.30  &  79.97  &  3.64\\
2020-10-29 & 79.00  &  78.83  &  76.16  &  0.25\\
2020-10-30 & 80.70  &  78.25  &  80.65  &  0\\
\hline
    \end{tabular}
    
    \label{table:SH600309}
\end{table}

\begin{table}
    \centering
    \caption{SH600346 daily return}
    \begin{tabular}{c|cccc}
\hline
Date & Price at 11:20 & Real price at 14:50 & Predicted price at 14:50 & Daily return (\%)\\
\hline
2020-10-9  & 19.05  &  19.01  &  18.90  &  0.22\\
2020-10-12 & 19.84  &  19.89  &  20.30  &  0.27\\
2020-10-13 & 20.47  &  20.88  &  21.10  &  2.21\\
2020-10-14 & 20.87  &  21.07  &  20.90  &  1.08\\
2020-10-15 & 20.79  &  20.57  &  20.74  &  1.19\\
2020-10-16 & 20.42  &  20.37  &  20.41  &  0\\
2020-10-19 & 20.52  &  20.38  &  19.55  &  0.75\\
2020-10-20 & 20.21  &  20.49  &  20.26  &  1.51\\
2020-10-21 & 20.30  &  20.09  &  20.30  &  0\\
2020-10-22 & 19.33  &  19.48  &  18.70  &  -0.81\\
2020-10-23 & 19.66  &  19.50  &  19.56  &  0.86\\
2020-10-26 & 19.53  &  19.46  &  20.11  &  -0.38\\
2020-10-27 & 19.04  &  19.20  &  18.29  &  -0.86\\
2020-10-28 & 19.52  &  20.34  &  19.53  &  0\\
2020-10-29 & 20.01  &  19.99  &  20.00  &  0\\
2020-10-30 & 19.62  &  19.49  &  19.56  &  0.70\\
\hline
    \end{tabular}
    
    \label{table:SH600346}
\end{table}

\begin{table}
    \centering
    \caption{SH600519 daily return}
    \begin{tabular}{c|cccc}
\hline
Date & Price at 11:20 & Real price at 14:50 & Predicted price at 14:50 & Daily return (\%)\\
\hline
2020-10-9  & 1703.75  &  1696.41  &  1703.15  &  0\\
2020-10-12 & 1742.00  &  1748.24  &  1736.77  &  -0.37\\
2020-10-13 & 1741.31  &  1739.13  &  1741.28  &  0\\
2020-10-14 & 1730.13  &  1729.03  &  1729.02  &  0\\
2020-10-15 & 1732.50  &  1724.30  &  1732.14  &  0\\
2020-10-16 & 1722.99  &  1713.50  &  1730.24  &  -0.57\\
2020-10-19 & 1696.00  &  1697.68  &  1693.43  &  -0.10\\
2020-10-20 & 1729.90  &  1733.38  &  1730.82  &  0\\
2020-10-21 & 1727.52  &  1731.80  &  1727.46  &  0\\
2020-10-22 & 1731.85  &  1743.00  &  1729.72  &  -0.67\\
2020-10-23 & 1739.00  &  1722.99  &  1742.37  &  -0.96\\
2020-10-26 & 1645.84  &  1642.00  &  1649.04  &  -0.23\\
2020-10-27 & 1618.55  &  1625.31  &  1618.72  &  0\\
2020-10-28 & 1657.57  &  1665.29  &  1659.36  &  0.46\\
2020-10-29 & 1705.00  &  1688.87  &  1757.35  &  -0.97\\
2020-10-30 & 1688.98  &  1666.60  &  1688.38  &  0\\
\hline
    \end{tabular}
    
    \label{table:SH600519}
\end{table}

\begin{table}
    \centering
    \caption{SH601012 daily return}
    \begin{tabular}{c|cccc}
\hline
Date & Price at 11:20 & Real price at 14:50 & Predicted price at 14:50 & Daily return (\%)\\
\hline
2020-10-9  & 82.43  &  81.24  &  82.18  &  1.59\\
2020-10-12 & 82.51  &  82.75  &  82.07  &  -0.32\\
2020-10-13 & 81.85  &  82.54  &  81.72  &  -0.92\\
2020-10-14 & 80.46  &  80.74  &  81.99  &  0.37\\
2020-10-15 & 78.27  &  77.64  &  76.49  &  0.84\\
2020-10-16 & 76.37  &  76.02  &  76.87  &  -0.47\\
2020-10-19 & 72.00  &  71.09  &  69.06  &  1.21\\
2020-10-20 & 71.46  &  73.14  &  71.31  &  -2.24\\
2020-10-21 & 72.21  &  72.47  &  71.59  &  -0.35\\
2020-10-22 & 70.19  &  70.83  &  70.12  &  -0.85\\
2020-10-23 & 69.69  &  67.71  &  70.21  &  -2.64\\
2020-10-26 & 69.49  &  70.20  &  69.69  &  0.95\\
2020-10-27 & 69.46  &  71.38  &  69.44  &  0\\
2020-10-28 & 72.12  &  72.39  &  71.70  &  -0.36\\
2020-10-29 & 73.88  &  74.33  &  74.08  &  0.6\\
2020-10-30 & 75.88  &  74.80  &  75.32  &  1.44\\
\hline
    \end{tabular}
    
    \label{table:SH601012}
\end{table}

\begin{table}
    \centering
    \caption{SH601088 daily return}
    \begin{tabular}{c|cccc}
\hline
Date & Price at 11:20 & Real price at 14:50 & Predicted price at 14:50 & Daily return (\%)\\
\hline
2020-10-9  & 16.74  &  16.56  &  16.76  &  -1.09\\
2020-10-12 & 16.80  &  16.73  &  16.80  &  0\\
2020-10-13 & 16.66  &  16.77  &  16.81  &  0.67\\
2020-10-14 & 16.59  &  16.66  &  16.58  &  0\\
2020-10-15 & 16.91  &  16.77  &  17.02  &  -0.85\\
2020-10-16 & 16.98  &  16.97  &  17.04  &  -0.06\\
2020-10-19 & 16.88  &  16.86  &  16.93  &  -0.12\\
2020-10-20 & 16.71  &  16.82  &  16.96  &  0.67\\
2020-10-21 & 16.73  &  16.76  &  16.74  &  0\\
2020-10-22 & 16.70  &  16.87  &  16.72  &  1.03\\
2020-10-23 & 16.67  &  16.71  &  16.70  &  0.24\\
2020-10-26 & 16.65  &  16.57  &  16.64  &  0\\
2020-10-27 & 16.77  &  16.75  &  17.05  &  -0.12\\
2020-10-28 & 16.57  &  16.64  &  16.59  &  0.43\\
2020-10-29 & 16.48  &  16.55  &  16.50  &  0.43\\
2020-10-30 & 16.81  &  16.63  &  16.84  &  -1.09\\
\hline
    \end{tabular}
    
    \label{table:SH601088}
\end{table}

\begin{table}
    \centering
    \caption{SH601225 daily return}
    \begin{tabular}{c|cccc}
\hline
Date & Price at 11:20 & Real price at 14:50 & Predicted price at 14:50 & Daily return (\%)\\
\hline
2020-10-9  & 8.96  &  8.89  &  9.33  &  -0.83\\
2020-10-12 & 8.99  &  8.99  &  8.98  &  0\\
2020-10-13 & 9.11  &  9.25  &  9.09  &  -1.67 \\
2020-10-14 & 9.09  &  9.13  &  9.10  &  0\\
2020-10-15 & 9.40  &  9.14  &  9.45  &  -3.10\\
2020-10-16 & 9.30  &  9.35  &  9.35  &  0.60\\
2020-10-19 & 9.28  &  9.17  &  9.25  &  1.31\\
2020-10-20 & 9.12  &  9.17  &  9.24  &  0.60\\
2020-10-21 & 9.03  &  9.08  &  9.04  &  0.60\\
2020-10-22 & 8.92  &  9.01  &  8.94  &  1.07\\
2020-10-23 & 8.99  &  8.90  &  8.98  &  1.07\\
2020-10-26 & 8.83  &  8.91  &  8.82  &  -0.95\\
2020-10-27 & 8.83  &  8.88  &  8.85  &  0.60\\
2020-10-28 & 8.81  &  8.92  &  8.82  &  0\\
2020-10-29 & 8.74  &  8.79  &  8.74  &  0\\
2020-10-30 & 8.90  &  8.74  &  8.91  &  0\\
\hline
    \end{tabular}
    
    \label{table:SH601225}
\end{table}

\begin{table}
    \centering
    \caption{SH601318 daily return}
    \begin{tabular}{c|cccc}
\hline
Date & Price at 11:20 & Real price at 14:50 & Predicted price at 14:50 & Daily return (\%)\\
\hline
2020-10-9  & 78.41  &  77.88  &  78.92  &  -0.69\\
2020-10-12 & 80.38  &  80.96  &  80.01  &  -0.76\\
2020-10-13 & 81.11  &  81.46  &  82.15  &  0.46 \\
2020-10-14 & 81.29  &  81.35  &  81.29  &  0\\
2020-10-15 & 82.12  &  81.50  &  80.97  &  0.81\\
2020-10-16 & 82.16  &  82.13  &  82.28  &  -0.04\\
2020-10-19 & 82.73  &  81.91  &  81.98  &  1.08\\
2020-10-20 & 81.19  &  81.26  &  81.31  &  0.09\\
2020-10-21 & 81.98  &  82.50  &  82.02  &  0\\
2020-10-22 & 81.35  &  82.36  &  81.25  &  -1.32\\
2020-10-23 & 84.51  &  83.75  &  84.79  &  -1.00\\
2020-10-26 & 82.88  &  82.10  &  83.24  &  -1.02\\
2020-10-27 & 81.10  &  81.29  &  81.12  &  0\\
2020-10-28 & 79.29  &  79.65  &  78.97  &  -0.47\\
2020-10-29 & 78.81  &  79.16  &  78.40  &  -0.46\\
2020-10-30 & 78.31  &  77.76  &  78.40  &  -0.72\\
\hline
    \end{tabular}
    
    \label{table:SH601318}
\end{table}

\begin{table}
    \centering
    \caption{SH601398 daily return}
    \begin{tabular}{c|cccc}
\hline
Date & Price at 11:20 & Real price at 14:50 & Predicted price at 14:50 & Daily return (\%)\\
\hline
2020-10-9  & 4.91  &  4.92  &  4.93  &  0.20\\
2020-10-12 & 4.94  &  4.96  &  5.02  &  0.41\\
2020-10-13 & 4.93  &  4.92  &  4.99  &  -0.20 \\
2020-10-14 & 4.92  &  4.92  &  4.92  &  0\\
2020-10-15 & 4.95  &  4.93  &  4.94  &  0.41\\
2020-10-16 & 5.00  &  4.99  &  5.08  &  -0.20\\
2020-10-19 & 5.03  &  5.02  &  5.03  &  0\\
2020-10-20 & 4.96  &  4.97  &  4.97  &  0.20\\
2020-10-21 & 5.02  &  5.05  &  5.02  &  0\\
2020-10-22 & 5.04  &  5.05  &  5.04  &  0\\
2020-10-23 & 5.07  &  5.07  &  5.08  &  0\\
2020-10-26 & 5.04  &  5.02  &  5.05  &  -0.41\\
2020-10-27 & 5.02  &  5.03  &  5.02  &  0\\
2020-10-28 & 5.00  &  5.01  &  5.00  &  0\\
2020-10-29 & 4.98  &  4.97  &  4.93  &  0.20\\
2020-10-30 & 4.93  &  4.92  &  4.93  &  0\\
\hline
    \end{tabular}
    
    \label{table:SH601398}
\end{table}

\begin{table}
    \centering
    \caption{SH601766 daily return}
    \begin{tabular}{c|cccc}
\hline
Date & Price at 11:20 & Real price at 14:50 & Predicted price at 14:50 & Daily return (\%)\\
\hline
2020-10-9  & 5.56  &  5.56  &  5.56  &  0\\
2020-10-12 & 5.64  &  5.65  &  5.63  &  -0.18\\
2020-10-13 & 5.62  &  5.61  &  5.61  &  0.18 \\
2020-10-14 & 5.58  &  5.58  &  5.67  &  0\\
2020-10-15 & 5.56  &  5.55  &  5.57  &  -0.18\\
2020-10-16 & 5.57  &  5.59  &  5.58  &  0.36\\
2020-10-19 & 5.59  &  5.57  &  5.46  &  0.36\\
2020-10-20 & 5.54  &  5.55  &  5.53  &  -0.18\\
2020-10-21 & 5.52  &  5.60  &  5.52  &  0\\
2020-10-22 & 5.55  &  5.54  &  5.53  &  0.18\\
2020-10-23 & 5.56  &  5.56  &  5.55  &  0\\
2020-10-26 & 5.56  &  5.54  &  5.56  &  0\\
2020-10-27 & 5.52  &  5.52  &  5.53  &  0\\
2020-10-28 & 5.49  &  5.51  &  5.49  &  0\\
2020-10-29 & 5.48  &  5.46  &  5.46  &  0.36\\
2020-10-30 & 5.46  &  5.40  &  5.47  &  -1.09\\
\hline
    \end{tabular}
    
    \label{table:SH601766}
\end{table}

\begin{table}
    \centering
    \caption{SH601857 daily return}
    \begin{tabular}{c|cccc}
\hline
Date & Price at 11:20 & Real price at 14:50 & Predicted price at 14:50 & Daily return (\%)\\
\hline
2020-10-9  & 4.13  &  4.14  &  4.13  &  0\\
2020-10-12 & 4.16  &  4.17  &  4.23  &  0.24\\
2020-10-13 & 4.15  &  4.14  &  4.14  &  0.24 \\
2020-10-14 & 4.11  &  4.11  &  4.07  &  0\\
2020-10-15 & 4.11  &  4.10  &  4.11  &  0\\
2020-10-16 & 4.11  &  4.12  &  4.12  &  0.24\\
2020-10-19 & 4.11  &  4.11  &  4.18  &  0\\
2020-10-20 & 4.09  &  4.09  &  4.09  &  0\\
2020-10-21 & 4.09  &  4.11  &  4.09  &  0\\
2020-10-22 & 4.09  &  4.09  &  4.09  &  0\\
2020-10-23 & 4.12  &  4.13  &  4.14  &  0.24\\
2020-10-26 & 4.09  &  4.11  &  4.10  &  0.49\\
2020-10-27 & 4.10  &  4.09  &  4.09  &  0.24\\
2020-10-28 & 4.09  &  4.09  &  4.08  &  0\\
2020-10-29 & 4.06  &  4.07  &  4.06  &  0\\
2020-10-30 & 4.10  &  4.08  &  4.10  &  0\\
\hline
    \end{tabular}
    
    \label{table:SH601857}
\end{table}

\begin{table}
    \centering
    \caption{SH601888 daily return}
    \begin{tabular}{c|cccc}
\hline
Date & Price at 11:20 & Real price at 14:50 & Predicted price at 14:50 & Daily return (\%)\\
\hline
2020-10-9  & 201.10  &  204.84  &  198.49  &  -1.689\\
2020-10-12 & 208.30  &  211.60  &  215.31  &  1.48\\
2020-10-13 & 209.90  &  214.79  &  209.15  &  -2.19 \\
2020-10-14 & 208.11  &  209.77  &  208.28  &  0\\
2020-10-15 & 202.91  &  202.48  &  202.72  &  0\\
2020-10-16 & 202.72  &  203.40  &  202.91  &  0\\
2020-10-19 & 200.01  &  199.99  &  200.40  &  -0.01\\
2020-10-20 & 200.59  &  201.16  &  198.40  &  -0.26\\
2020-10-21 & 195.71  &  195.47  &  195.37  &  0.11\\
2020-10-22 & 191.53  &  195.93  &  192.37  &  1.97\\
2020-10-23 & 193.41  &  194.30  &  189.56  &  -0.4\\
2020-10-26 & 185.81  &  184.07  &  186.45  &  -0.78\\
2020-10-27 & 184.76  &  184.74  &  184.15  &  0.01\\
2020-10-28 & 197.18  &  199.79  &  197.33  &  0\\
2020-10-29 & 202.81  &  201.80  &  210.94  &  -0.45\\
2020-10-30 & 203.34  &  200.05  &  203.04  &  1.48\\
\hline
    \end{tabular}
    
    \label{table:SH601888}
\end{table}

\begin{table}
    \centering
    \caption{SH601899 daily return}
    \begin{tabular}{c|cccc}
\hline
Date & Price at 11:20 & Real price at 14:50 & Predicted price at 14:50 & Daily return (\%)\\
\hline
2020-10-9  & 6.24  &  6.21  &  6.23  &  0.49\\
2020-10-12 & 6.54  &  6.56  &  6.77  &  0.33\\
2020-10-13 & 6.50  &  6.59  &  6.49  &  -1.46 \\
2020-10-14 & 6.52  &  6.49  &  6.51  &  0.49\\
2020-10-15 & 6.57  &  6.56  &  6.58  &  0\\
2020-10-16 & 6.62  &  6.60  &  6.65  &  -0.33\\
2020-10-19 & 6.61  &  6.67  &  6.64  &  0.98\\
2020-10-20 & 6.72  &  6.77  &  6.68  &  -0.81\\
2020-10-21 & 6.91  &  6.95  &  6.98  &  0.65\\
2020-10-22 & 6.91  &  6.96  &  6.90  &  0\\
2020-10-23 & 6.96  &  6.87  &  6.97  &  -1.46\\
2020-10-26 & 7.00  &  7.03  &  7.03  &  0.49\\
2020-10-27 & 7.01  &  7.06  &  7.00  &  0\\
2020-10-28 & 7.06  &  7.16  &  7.06  &  0\\
2020-10-29 & 7.05  &  7.02  &  7.17  &  -0.49\\
2020-10-30 & 7.16  &  6.98  &  7.16  &  0\\
\hline
    \end{tabular}
    
    \label{table:SH601899}
\end{table}

\begin{table}
    \centering
    \caption{SH601939 daily return}
    \begin{tabular}{c|cccc}
\hline
Date & Price at 11:20 & Real price at 14:50 & Predicted price at 14:50 & Daily return (\%)\\
\hline
2020-10-9  & 6.14  &  6.13  &  6.14  &  0\\
2020-10-12 & 6.18  &  6.23  &  6.17  &  -0.81\\
2020-10-13 & 6.19  &  6.17  &  6.19  &  0 \\
2020-10-14 & 6.17  &  6.16  &  6.16  &  0\\
2020-10-15 & 6.23  &  6.22  &  6.23  &  0\\
2020-10-16 & 6.40  &  6.40  &  6.38  &  0\\
2020-10-19 & 6.57  &  6.51  &  6.64  &  -0.98\\
2020-10-20 & 6.34  &  6.35  &  6.10  &  -0.16\\
2020-10-21 & 6.38  &  6.43  &  6.38  &  0\\
2020-10-22 & 6.42  &  6.46  &  6.42  &  0\\
2020-10-23 & 6.50  &  6.49  &  6.51  &  -0.16\\
2020-10-26 & 6.44  &  6.45  &  6.43  &  -0.16\\
2020-10-27 & 6.39  &  6.38  &  6.40  &  -0.16\\
2020-10-28 & 6.30  &  6.28  &  6.31  &  -0.33\\
2020-10-29 & 6.30  &  6.30  &  6.28  &  0\\
2020-10-30 & 6.33  &  6.27  &  6.28  &  0.98\\
\hline
    \end{tabular}
    
    \label{table:SH601939}
\end{table}

\begin{table}
    \centering
    \caption{SH603288 daily return}
    \begin{tabular}{c|cccc}
\hline
Date & Price at 11:20 & Real price at 14:50 & Predicted price at 14:50 & Daily return (\%)\\
\hline
2020-10-9  & 164.56  &  162.62  &  163.56  &  1.20\\
2020-10-12 & 171.30  &  172.28  &  182.68  &  0.60\\
2020-10-13 & 173.20  &  172.03  &  176.88  &  -0.72 \\
2020-10-14 & 171.25  &  171.80  &  171.43  &  0.34\\
2020-10-15 & 169.36  &  170.87  &  168.93  &  -0.93\\
2020-10-16 & 168.27  &  168.41  &  168.82  &  0.09\\
2020-10-19 & 164.46  &  163.13  &  163.12  &  0.82\\
2020-10-20 & 167.71  &  169.59  &  172.16  &  1.16\\
2020-10-21 & 168.60  &  168.99  &  169.20  &  0.24\\
2020-10-22 & 167.18  &  168.20  &  167.24  &  0\\
2020-10-23 & 166.50  &  164.96  &  166.40  &  0\\
2020-10-26 & 161.49  &  162.00  &  161.92  &  0.31\\
2020-10-27 & 163.84  &  164.23  &  163.63  &  -0.24\\
2020-10-28 & 165.89  &  166.00  &  167.59  &  0.07\\
2020-10-29 & 167.90  &  168.02  &  167.59  &  -0.07\\
2020-10-30 & 162.45  &  159.45  &  157.43  &  1.85\\
\hline
    \end{tabular}
    
    \label{table:SH603288}
\end{table}
\end{document}